\documentclass[twocolumn,nobalancelastpage,superscriptaddress,amsmath,amssymb,oneside,aps,prd,10pt]{revtex4-1}
\usepackage[english]{babel}
\usepackage[utf8]{inputenc}
\usepackage[sc]{mathpazo}
\usepackage{url}
\usepackage{xspace}
\usepackage{graphicx}
\usepackage{siunitx}
\usepackage{xcolor}
\usepackage{lmodern}

\definecolor{darkred}{rgb}{0.5,0,0}
\definecolor{darkblue}{rgb}{0,0,0.5}
\definecolor{firebrick}{rgb}{0.75,0.125,0.125}
\definecolor{darkgreen}{rgb}{0,0.5,0}
\definecolor{darkyellow}{rgb}{0.5,0.5,0}
\definecolor{darkcyan}{rgb}{0,0.5,0.5}
\definecolor{rulecolor}{gray}{0.8}
\usepackage[colorlinks=true,linkcolor=firebrick,citecolor=darkgreen,urlcolor=darkblue]{hyperref}

\newcommand{\tb}[1]{Table\,\ref{tab:#1}}
\newcommand{\fg}[1]{Fig.\,\ref{fig:#1}}
\newcommand{\eq}[1]{Eq.\,\ref{eq:#1}}
\newcommand{\dd}{\mathrm{d}}

\DeclareMathOperator{\ex}{E}
\DeclareMathOperator{\var}{Var}
\newcommand{\avg}[1]{\langle{#1}\rangle}

\newcommand{\mlna}{\avg{\ln A}}

\newcommand{\mlnn}{\avg{\ln N_\mu}}
\newcommand{\lnmn}{\ln \avg{N_\mu}}
\newcommand{\mlnhn}{\avg{\ln \hat N_\mu}}
\newcommand{\bigo}{\mathcal O}
\newcommand{\norm}{\mathcal N}
\begin{document}

\title{Computing mean logarithmic mass from muon counts in air shower experiments}

\author{H.P.~Dembinski}
\email[Corresponding author: ]{hdembins@mpi-hd.mpg.de}
\affiliation{Bartol Institute, University of Delaware \& Max-Planck-Institute for Nuclear Physics}

\begin{abstract}
I discuss the conversion of muon counts in air showers, which are observable by experiments, into mean logarithmic mass, an important variable to express the mass composition of cosmic rays. Stochastic fluctuations in the shower development and statistical fluctuations from muon sampling can subtly bias the conversion. A central theme is that the mean of the logarithm of the muon number is not identical to the logarithm of the mean. It is discussed how that affects the conversion in practice. Simple analytical formulas to quantify and correct such biases are presented, which are applicable to any kind of experiment.
\end{abstract}

\maketitle

\section{Introduction}

The mean logarithmic mass $\mlna$ is a common variable to summarize the mass composition of cosmic rays. Most ground-based experiments infer the mass by counting muons in cosmic-ray induced air showers~\cite{Kampert_cr_review}. This paper discusses the conversion of muon number to mean logarithmic mass from the point of view of the data analyst, with a focus on the effect of stochastic fluctuations in the shower development and the detector response on the conversion. The fluctuations can bias estimates of $\mlna$ in several ways. Biases here are defined in the usual statistical sense; if $\hat x$ is an estimate of the true value $x$ that fluctuates according to a probability density $f(\hat x)$, then the bias is the expectation $\ex[\hat x - x] = \int (\hat x - x) f(\hat x) \text{d}\hat x$. We generally want $\hat x$ to have zero bias, so that the sample average $\avg{\hat x}$ converges to $x$ for large samples.

The results in this paper are not specific to a particular type of experiment. It is assumed throughout this paper that an experiment provides an unbiased estimate $\hat N_\mu$ of the total number of muons $N_\mu$ produced in an air shower and an estimate $\hat E$ of the shower energy $E$. This is far from trivial and much of the difficulty in running an experiment deals with this. The total number of muons $N_\mu$ produced in an air shower cannot be directly measured, because experiments can only count muons that reach the ground, while some decay on the way. The experimental distinction between muons and other charged particles at the ground is not easy either~\cite{AugerInclined,IceTop,KASCADE,HiresMia}. But in principle, $\hat N_\mu$ can be inferred for a given geometry and shower energy from the measurement by applying an average correction obtained from air shower simulations. Highly-inclined air showers recorded by Haverah Park and the Pierre Auger Observatory have been analyzed in this way~\cite{Ave:2000xs,Ave:2000dd,Dembinski:2009jc,Aab:2014pza}. Similarly, an estimate $\hat E$ of the shower energy can be inferred from the number of electrons and photons that reach the ground, or by recording the longitudinal shower profile with telescopes.

The paper deals with the comparably easier part of the conversion of the unbiased estimates $\hat E, \hat N_\mu$ to $\mlna$. Fluctuations occur in the shower development and arise from the sampling of an air shower by a detector. It is important to distinguish between these two kinds of fluctuations, because they are approximately independent~\cite{Dembinski:2015wqa}. Both randomly shift the estimates $\hat E, \hat N_\mu$ away from their true values $E,N_\mu$, and these random shifts cause some subtle biases in the conversion to $\mlna$. We quantify these biases. Knowing their sizes allows one to safely neglected them if they are small, and to correct them otherwise.

\section{From muon number to mass}\label{sec:from_muon_to_mass}

It is instructive to introduce fluctuations step-by-step. We start by ignoring fluctuations from detector sampling and consider only stochastic fluctuations in the shower development. The true muon number $N_\mu$ and the shower energy $E$ shall be exactly known and the energy $E$ shall be same for all showers. Stochastic fluctuations in the hadronic interactions are still causing the muon number $N_\mu$ to vary randomly.

The first point to make is that $\mlna$ is best computed from the mean logarithmic muon number $\mlnn$, and not the mean of the muon number $\avg{N_\mu}$. In either case, the average here is computed over many air showers with the same shower energy $E$.

The following argument is similar to the one developed by the Pierre Auger collaboration for the depth of shower maximum~\cite{Abreu:2013env}. The relationship between $N_\mu$ and $A$ can be understood within the Matthews-Heitler model of a hadronic shower~\cite{Matthews2005}. The analytical model treats air showers in a simplified way, but describes surprisingly many features of air showers correctly. According to the model, the total muon number $N_\mu$ for a cosmic ray with $A$ nucleons scales with a power of the number of nucleons
\begin{equation}
N_\mu = A^{1 - \beta} \, N_{\mu}^p,
\label{eq:nmua}
\end{equation}
where $N_\mu^p$ is the number of muons in a proton-induced air shower, and $\beta \simeq 0.9$ is a constant.

\begin{figure}
\vspace{-0.4cm}
\includegraphics[width=\columnwidth]{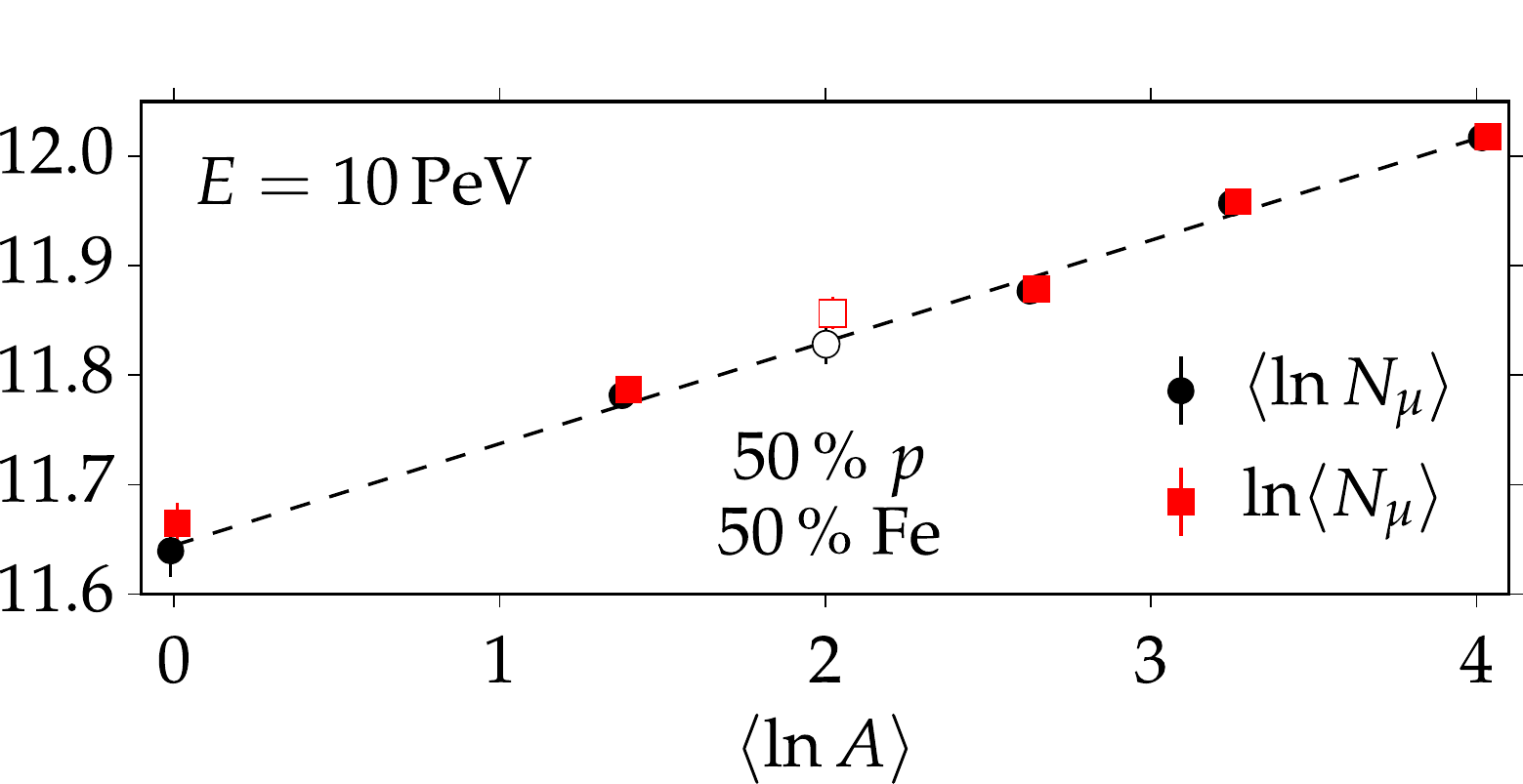}
\includegraphics[width=\columnwidth]{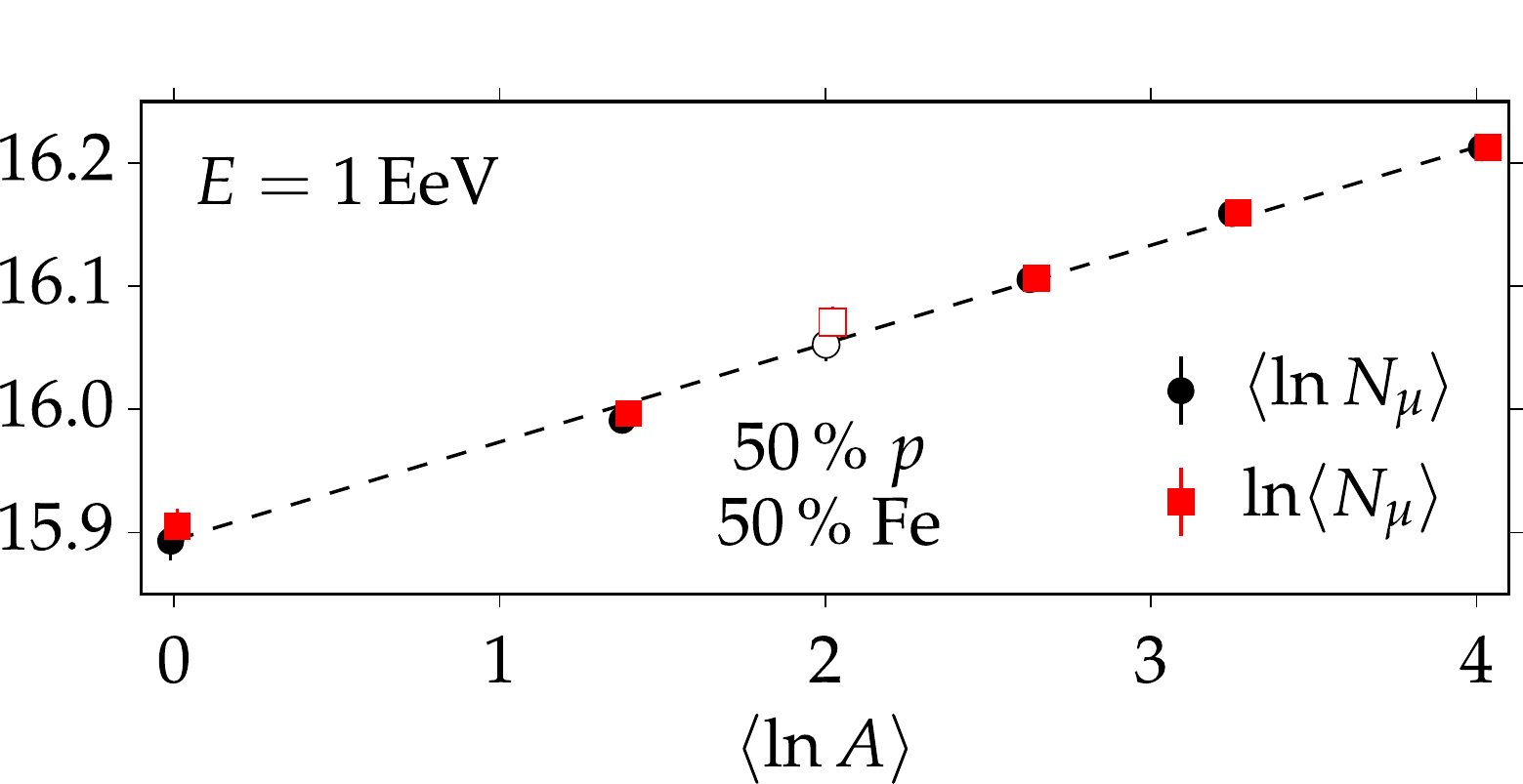}
\includegraphics[width=\columnwidth]{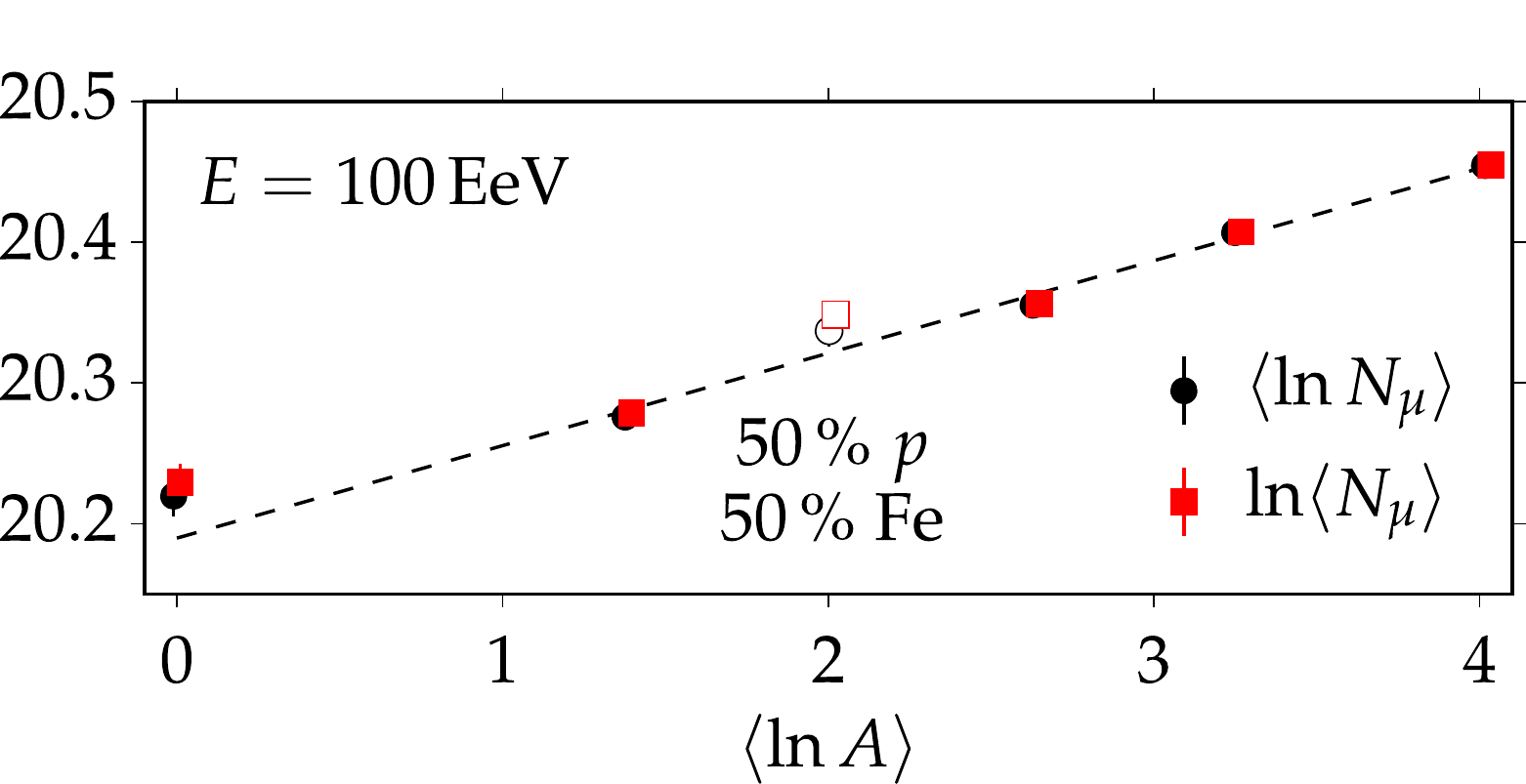}
\caption{Average logarithm of the number of muons $\mlnn$ (circles) and logarithm of the average number of muons $\lnmn$ (squares) in simulated vertical air showers produced by primary particles with $A$ nucleons. A fitted straight line (dashed) is shown for comparison. Solid markers stand for averages computed over showers from a single primary, open markers stand for an equal mix of proton and iron showers. Error bars indicate the statistical uncertainty of the finite sample (2200 showers per primary and energy).}
\label{fig:lnn_vs_lna}
\end{figure}

This behavior is well reproduced in full air shower simulations. In the Matthews-Heitler model, stochastic fluctuations in the shower development are neglected. To show that \eq{nmua} holds for the real showers, several sets of vertical showers with identical primary particles were simulated with CORSIKA~\cite{CORSIKA} compiled with the CONEX option, using the hadronic interaction models SIBYLL-2.3~\cite{SIBYLL23} and GHEISHA~\cite{GHEISHA}. The showers were simulated in a US standard atmosphere until a slant depth of \SI{1050}{g cm^{-2}}. The number of muons $N_\mu$ in each shower were taken from the maximum of the longitudinal muon profile. Proton, helium, nitrogen, silicon, and iron primaries were simulated. For each primary, the averages $\lnmn$ and $\mlnn$ were computed. The two are subtly different, because the expectation is noncommutative with a non-linear mapping $f(x)$, $\ex[f(x)] \neq f(\ex[x])$. The dependence on $A$ is shown in \fg{lnn_vs_lna} for a wide range of primary energies. For a pure composition where all showers are initated by a primary with the same mass $A$, both $\lnmn$ and $\mlnn$ scale with $\ln A$ as predicted by \eq{nmua}. This result is independent of the hadronic interaction model and shower inclination.

To use \eq{nmua} to get an estimate of $\mlna$ for real air showers, we consider the realistic case where the mass $A$ is another stochastic variable that changes from shower to shower. For a pure composition, the simulations showed that $\avg{N_\mu} = A^{1-\beta}\,\avg{N_\mu^p}$ . If $f_A$ is the fraction of primaries with $A$ nucleons in a mixed composition, we have
\begin{equation}
\begin{gathered}
\sum_A f_A \avg{N_\mu} = \sum_A f_A \, A^{1-\beta} \, \avg{N_\mu^p} = \avg{N_\mu^p} \sum_A f_A \, A^{1-\beta} \\
\Leftrightarrow \quad \avg{N_\mu} = \avg{N_\mu^p} \, \avg{A^{1-\beta}}.
\end{gathered}
\end{equation}
Unfortunately, we cannot convert $\avg{A^{1-\beta}}$ to $\avg{A}$ or $\mlna$, because these are non-linear functions of $A$. The solution is to start from $\mlnn = (1-\beta) \ln A + \avg{\ln N_\mu^p}$ for a pure composition, which is also supported by the simulations. Then the result of the superposition is
\begin{equation}
\mlnn = (1 - \beta) \mlna + \avg{\ln N_\mu^p},
\label{eq:mlnn}
\end{equation}
where we used that $\avg{a x + b y} = a \avg{x} + b \avg{y}$ for constants $a,b$ and stochastic variables $x,y$.

Both $\beta$ and $\avg{\ln N_\mu^p}$ can be obtained from air shower simulations. If $\avg{\ln N_\mu^\text{Fe}}$ is available, it can be used to substitute $\beta$. The two related formulas for $\mlna$ are
\begin{align}
\mlna &= \frac{\mlnn - \avg{\ln N_\mu^p}}{1 - \beta}
\label{eq:mlna_1} \\
\mlna &= \frac{\mlnn - \avg{\ln N_\mu^p}}{\avg{\ln N_\mu^\text{Fe}} - \avg{\ln N_\mu^p}} \ln 56.
\label{eq:mlna_2}
\end{align}
This approach is very elegant, because the equations are true whatever the probability distributions are for $A$, $N_\mu$, $N_\mu^p$, and $N_\mu^\text{Fe}$.

As previously stated, the mean of the logarithm is not the same as the logarithm of the mean, $\lnmn$ is always higher than $\mlnn$. Still, the two are quite close and the bias of substituting one for the other may be negligible in some situations. To judge when this is safe, a simple formula to compute the bias is given in section~\ref{sec:lnmn_to_mlnn}. Some analyses~\cite{Dembinski2015_surface_muons} do not produce an estimate of the muon number event-by-event, only the average $\avg{N_\mu}$ over many showers. In these cases, the formula can be used to correct the difference $(\mlnn-\lnmn)$.

So far fluctuations introduced by detector sampling were neglected, but $N_\mu$ is not known in practice, only an estimate $\hat N_\mu$ which fluctuates around $N_\mu$. Some muons decay on the way to the ground, the detector does not count all muons that arrive, and so on. It is assumed that these losses are corrected on average, but they introduces additional fluctuations. Since the mean of the logarithm is not the logarithm of the mean, we find $\mlnhn \neq \avg{\ln N_\mu}$ even if $\hat N_\mu$ is an unbiased estimate of $N_\mu$. How to correct for this effect is discussed in section~\ref{sec:detector_response}.

Finally, one has to consider that the average $\mlnhn$ is not computed over showers with the same energy $E$ in practice, but for showers that fall into the same energy bin. The energy $E$ is also not known exactly, only an estimate $\hat E$ of it. The quantitative impact of that is calculated in section~\ref{sec:energy_bias}.

\section{Muon number: Mean logarithm and logarithm of mean}\label{sec:lnmn_to_mlnn}

The difference $(\mlnn-\lnmn)$ can be calculated with a simple formula. To derive it, we use the following general substitution
\begin{equation}
N_\mu = \langle N_\mu \rangle (1 + \epsilon),
\label{eq:nmu_as_eps}
\end{equation}
where $\epsilon = (N_\mu - \langle N_\mu \rangle) / \langle N_\mu \rangle$ is the relative random deviation of the muon number from its mean. By construction, $\langle \epsilon \rangle = 0$. The average logarithmic muon number is
\begin{align}
\langle \ln N_\mu \rangle &= \langle \ln[\langle N_\mu \rangle (1 + \epsilon)] \rangle \nonumber \\
&= \ln \langle N_\mu \rangle + \langle \ln (1 + \epsilon) \rangle.
\end{align}
For small relative fluctuations, $\epsilon \ll 1$, the second logarithm can be expanded into a Taylor series,
\begin{align}
\langle \ln N_\mu \rangle &= \ln \langle N_\mu \rangle + \langle \epsilon - \frac 12  \epsilon^2 + \bigo(\epsilon^3) \rangle \nonumber \\
&= \ln \langle N_\mu \rangle - \frac 12 \langle \epsilon^2 \rangle + \langle \bigo( \epsilon^3 )\rangle.
\end{align}
The second-order term $\avg{\epsilon^2}$ is equal to the variance of the relative deviations from the mean,
\begin{align}
\avg{\epsilon^2} &= \avg{\epsilon^2} - \avg{\epsilon}^2 = \var[\epsilon] \nonumber \\
&= \var[(N_\mu - \langle N_\mu \rangle) / \langle N_\mu \rangle] = \var[N_\mu]/\langle N_\mu \rangle^2.
\label{eq:lnnvar}
\end{align}
Therefore, the offset can be computed for $\epsilon \ll 1$ as
\begin{equation}
\mlnn - \lnmn \approx - \frac 12 \var[(N_\mu - \langle N_\mu \rangle) / \langle N_\mu \rangle].
\label{eq:mlnn_from_lnmn}
\end{equation}

\begin{table}
\begin{tabular}{r c c c c c c}
$E / \mathrm{eV}$ & $10^{15}$ & $10^{16}$ & $10^{17}$ & $10^{18}$ & $10^{19}$ & $10^{20}$ \\
\hline
 & \multicolumn{6}{c}{p} \\
$\langle \ln N_\mu \rangle$ & $9.55$ & $11.64$ & $13.73$ & $15.89$ & $18.01$ & $20.22$\\
$\ln \langle N_\mu \rangle$ & $9.58$ & $11.66$ & $13.76$ & $15.91$ & $18.02$ & $20.23$\\
$\mathrm{Var}[\epsilon]$ & $0.051$ & $0.038$ & $0.038$ & $0.022$ & $0.030$ & $0.017$\\
 & \multicolumn{6}{c}{He} \\
$\langle \ln N_\mu \rangle$ & $9.66$ & $11.78$ & $13.890$ & $15.99$ & $18.143$ & $20.276$\\
$\ln \langle N_\mu \rangle$ & $9.67$ & $11.787$ & $13.894$ & $16.00$ & $18.145$ & $20.279$\\
$\mathrm{Var}[\epsilon]$ & $0.015$ & $0.010$ & $0.008$ & $0.011$ & $0.0049$ & $0.0057$\\
 & \multicolumn{6}{c}{N} \\
$\langle \ln N_\mu \rangle$ & $9.783$ & $11.877$ & $13.992$ & $16.105$ & $18.221$ & $20.355$\\
$\ln \langle N_\mu \rangle$ & $9.787$ & $11.879$ & $13.994$ & $16.107$ & $18.222$ & $20.356$\\
$\mathrm{Var}[\epsilon]$ & $0.008$ & $0.0042$ & $0.0035$ & $0.0027$ & $0.0027$ & $0.0019$\\
 & \multicolumn{6}{c}{Si} \\
$\langle \ln N_\mu \rangle$ & $9.887$ & $11.957$ & $14.048$ & $16.159$ & $18.274$ & $20.407$\\
$\ln \langle N_\mu \rangle$ & $9.891$ & $11.959$ & $14.049$ & $16.160$ & $18.275$ & $20.407$\\
$\mathrm{Var}[\epsilon]$ & $0.0067$ & $0.0029$ & $0.0021$ & $0.0015$ & $0.0012$ & $0.0010$\\
 & \multicolumn{6}{c}{Fe} \\
$\langle \ln N_\mu \rangle$ & $9.947$ & $12.017$ & $14.109$ & $16.212$ & $18.330$ & $20.454$\\
$\ln \langle N_\mu \rangle$ & $9.950$ & $12.018$ & $14.109$ & $16.213$ & $18.331$ & $20.455$\\
$\mathrm{Var}[\epsilon]$ & $0.0044$ & $0.0016$ & $0.0016$ & $0.0010$ & $0.00049$ & $0.00058$\\
\end{tabular}
\caption{Simulation results for vertical showers simulated with hadronic models SIBYLL2.3 and GHEISHA, as described in section \ref{sec:from_muon_to_mass}. The total number of muons $N_\mu$ is taken from the maximum of the longitudinal muon profile.}
\label{tab:nmu_sim}
\end{table}

\tb{nmu_sim} lists $\mlnn$, $\lnmn$, and $\var[(N_\mu - \langle N_\mu \rangle) / \langle N_\mu \rangle]$ for the air shower simulations described in the previous section. A useful empirical parametrization of the latter is shown in the appendix. The numbers confirm for single elements that $\var[\epsilon] \ll 1$, which implies $\epsilon \ll 1$. \eq{mlnn_from_lnmn} is therefore a good approximation for single primaries above $10^{15}\,\si{eV}$.

It also holds for any mix of primaries. The variance for a mix of primaries is larger than for a single primary, because the difference in the means $\langle N_\mu \rangle$ of different primaries contributes to the variance. With the data in \tb{nmu_sim}, $\var[\epsilon]$ was computed for all pairs of primaries. The largest value $\var[\epsilon] = 0.063$ is found at $10^{15}\,\si{eV}$ for a mix of proton and iron. This value is still small and thus \eq{mlnn_from_lnmn} remains valid.

% The standard deviation of $\epsilon$ depends on the mass composition. If several elements are mixed, $\epsilon$ also increases, because $\avg{N_\mu}_i$ differ for each species $i$. It can be directly read off \tb{nmu_sim}, I find that

% The relative variance $\var[(N_\mu - \langle N_\mu \rangle) / \langle N_\mu \rangle]$ of a data sample depends on its mass composition, which is not known in general. Therefore, experiments should focus on measuring $\mlnn$ directly. If only $\langle N_\mu \rangle$ can be measured, one should also measure $\var[(N_\mu - \langle N_\mu \rangle) / \langle N_\mu \rangle]$, and then apply \eq{mlnn_from_lnmn}. If the latter is also unknown, then it should be estimated from simulated air showers and the uncertainty propagated.

With these numbers, it is possible to address the question whether using $\lnmn$ instead of $\mlnn$ in \eq{mlna_1} or \ref{eq:mlna_2} introduces a noticeable bias. In the most extreme case, the bias is $(\lnmn - \mlnn) \approx 0.03$. In the conversion to $\mlna$, this bias is multiplied by a factor $1/(1 - \beta)$, see \eq{mlna_1}. For $\beta \simeq 0.9$, this is a factor of 10, so that the bias in $\mlna$ is $0.3$. This is about 7\,\% of the overall difference between proton and iron. Using the wrong mean makes the composition appear heavier than it truly is. The effect is small, but since the bias is easy to correct with \eq{mlnn_from_lnmn}, applying the correction is recommended.

\section{Bias from sampling fluctuations}\label{sec:detector_response}

The second type of difficulty in applying \eq{mlna_1} or \ref{eq:mlna_2} is that $N_\mu$ is not known, only an estimate $\hat N_\mu$. To measure $N_\mu$, an experiment would have to collect and count all muons with perfect accuracy. In reality, detectors sample only a small fraction of all particles, and cannot perfectly distinguish between muons and other shower particles. They measure an event-wise estimate $\hat N_\mu$ of $N_\mu$, which differs by a random offset for each shower.

This paper is only concerned with the effect of fluctuations, so it is again assumed that the estimate is unbiased, $\ex[\hat N_\mu] = N_\mu$. It still follows that $\mlnhn \neq \avg{\ln N_\mu}$, because of the fluctuations and the non-linear mapping.

A simple formula for the size of this bias can be derived analog to the previous section. The relative offset $\hat \epsilon = (\hat N_\mu - N_\mu) / N_\mu$ is introduced, which represents the additional random fluctuations introduced by the muon sampling. Typical values are again small, the Pierre Auger Observatory~\cite{DembinskiThesis,Aab:2014pza} achieves resolutions better than 30\,\%, so $\var[\hat\epsilon] < 0.09$. An expansion in a Taylor series for $\hat \epsilon \ll 1$ yields
\begin{align}
\mlnhn &= \langle \ln [N_\mu (1 + \hat \epsilon)] \rangle = \mlnn +  \langle \ln(1 + \hat \epsilon) \rangle \nonumber \\
&= \mlnn - \frac 12 \langle {\hat \epsilon}^2 \rangle + \bigo(\langle {\hat \epsilon}^3 \rangle).
\label{eq:detbias}
\end{align}
The term $\avg {\hat \epsilon}$ is zero, because $\hat{N}_\mu$ is unbiased. Values for $\avg{{\hat \epsilon}^2} = \var[(\hat N_\mu - N_\mu) / N_\mu]$ can be obtained from Monte-Carlo simulations of the experiment.

To give an example, the previously quoted value $\var[(\hat N_\mu - N_\mu) / N_\mu] = 0.09$ results in a bias $\avg{\ln \hat N_\mu} - \avg{\ln N_\mu} = -0.045$. Using \eq{mlna_1} and $\beta \simeq 0.9$, this translates into a bias in $\mlna$ of -0.45 or 11\,\% of the proton-iron distance, which makes the composition appear lighter.

\section{Bias from binning in energy}\label{sec:energy_bias}

In the previous sections, it was discussed how stochastic fluctuations of $N_\mu$ from shower-to-shower, and the additional fluctuations in its estimate $\hat N_\mu$ make it difficult to compute $\mlnn$, which is the natural quantity to convert to $\mlna$. It was assumed throughout that averages over $\ln N_\mu$ and $\ln \hat N_\mu$ can be computed for air showers with the exact same shower energy $E$, which is not possible in practice. In the final section, the bias from binning showers in energy is investigated, which is orthogonal to the effects discussed before. We will reach a point in complexity that cannot be handled with simple formulas anymore. The general case should be treated numerically or via a full Monte-Carlo simulation of the experiment.

Complexity is again introduced step-by-step. The true energy $E$ of each shower shall be known, but it now varies randomly from shower to shower. Showers then need to be binned in energy to compute an average of $\ln \hat N_\mu$, called $\mlnhn^\star$ for distinction. The offset $(\mlnhn - \mlnhn^\star)$ is investigated in the following.

Showers are sorted into a logarithmic energy interval $[\ln E_0, \ln E_1)$. The average $\mlnhn^\star$ is compared with the true value at the bin center $\avg{\ln E} = \frac 12 (\ln E_0 + \ln E_1)$. The cosmic ray flux has a steeply falling spectrum $\propto E^{-\gamma}$, therefore the event distribution inside the bin is very uneven, with more events near $\ln E_0$. This leads to a bias, since $\mlnn$ depends on the logarithm of the energy, $\mlnn = \beta \ln E + c$, where $c$ is a constant and the value of $\beta$ is very close to the one in \eq{mlna_1}, although they are not strictly the same. For the calculation, it does not matter whether they are exactly the same.

Lafferty and Wyatt~\cite{Lafferty1995} offered a general discussion of binning biases. As a remedy, they propose to adjust the horizontal placement of the data point in the bin. In general, it is simpler and equivalent to just compute the bias and correct for it. We will follow that strategy.

To compute the average value $\mlnhn^\star$ over an energy interval $\Delta \ln E = \ln E_1 - \ln E_0$, one has to integrate the argument over the interval weighted by the energy frequency $\propto E^{-\gamma}$. The result is expressed as a function of the expected value $\mlnhn = \ln N_\mu^0 + \beta \avg{\ln E}$. With $x = \ln E$, $\Delta x = x_1 - x_0$, and $E^{-\gamma} \dd E = e^{(1 - \gamma) x} \dd x$, I get
\begin{equation}
\mlnhn^\star = \frac{\int_{x_0}^{x_0 + \Delta x} (\ln N_\mu^0 + \beta x) e^{(1 - \gamma) x} \dd x}{\int_{x_0}^{x_0 + \Delta x} e^{(1 - \gamma) x} \dd x}.
\label{eq:binning_orig}
\end{equation}
For $\Delta x \ll 1$, I can use \eq{taylor3} from the appendix to approximate the result
\begin{align}
\mlnhn\!^\star &= \ln N_\mu^0 + \beta \left(x_0 + \frac {\Delta x} 2 + (1-\gamma) \frac{\Delta x^2}{12} \right) + \bigo(\Delta x^3) \nonumber \\
&= \mlnhn + \beta (1-\gamma) \frac{\Delta x^2}{12} + \bigo(\Delta x^3).
\end{align}
A typical bin width of 0.1 in $\log_{10}E$ is equivalent to $\Delta x \approx 0.23$, so that the higher orders can be neglected. With a spectral index $\gamma = 2.7$, and $\beta \simeq 0.9$, the bias for $\mlnhn$ is $-0.007$. This translates into a bias of $-0.07$ for $\mlna$, about 2\,\% of the proton-iron distance.

Alternatively, \eq{binning_orig} can be solved exactly by partial integration, but the resulting formula provides less insight. The point of this paper is to provide simple formulas to estimate the size of biases, therefore the Taylor expansion is shown here.

% I have so far assumed that $\mlna$ in \eq{mlnn} is constant. If the mass composition varies, then $\mlnn$ is not a linear function of $x$ anymore. Still, $\mlnn$ is typically slowly varying with $x$. It can therefore be well approximated by a Taylor series of low order in a narrow $x$-interval. \eq{taylor3} can be used to approximate the bias for arbitrary orders of such a series (exact results could also be obtained by successive partial integration).

Finally, one has to consider that the shower energy $E$ is also only known to a finite resolution. In practice, one only has an estimate $\hat E$ that varies stochastically around $E$. As before, it is assumed that $\hat E$ is an unbiased estimate for the energy. Events are sorted into energy bins based on the energy estimate $\hat E$, therefore also events with true energies outside of the bin interval contribute to the computation of $\mlnhn^\star$. Correcting for this effect is conceptually related to the \emph{unfolding} of resolution effects from distributions~\cite{Dembinski:2013hdz}.

\begin{figure}
\includegraphics[width=\columnwidth]{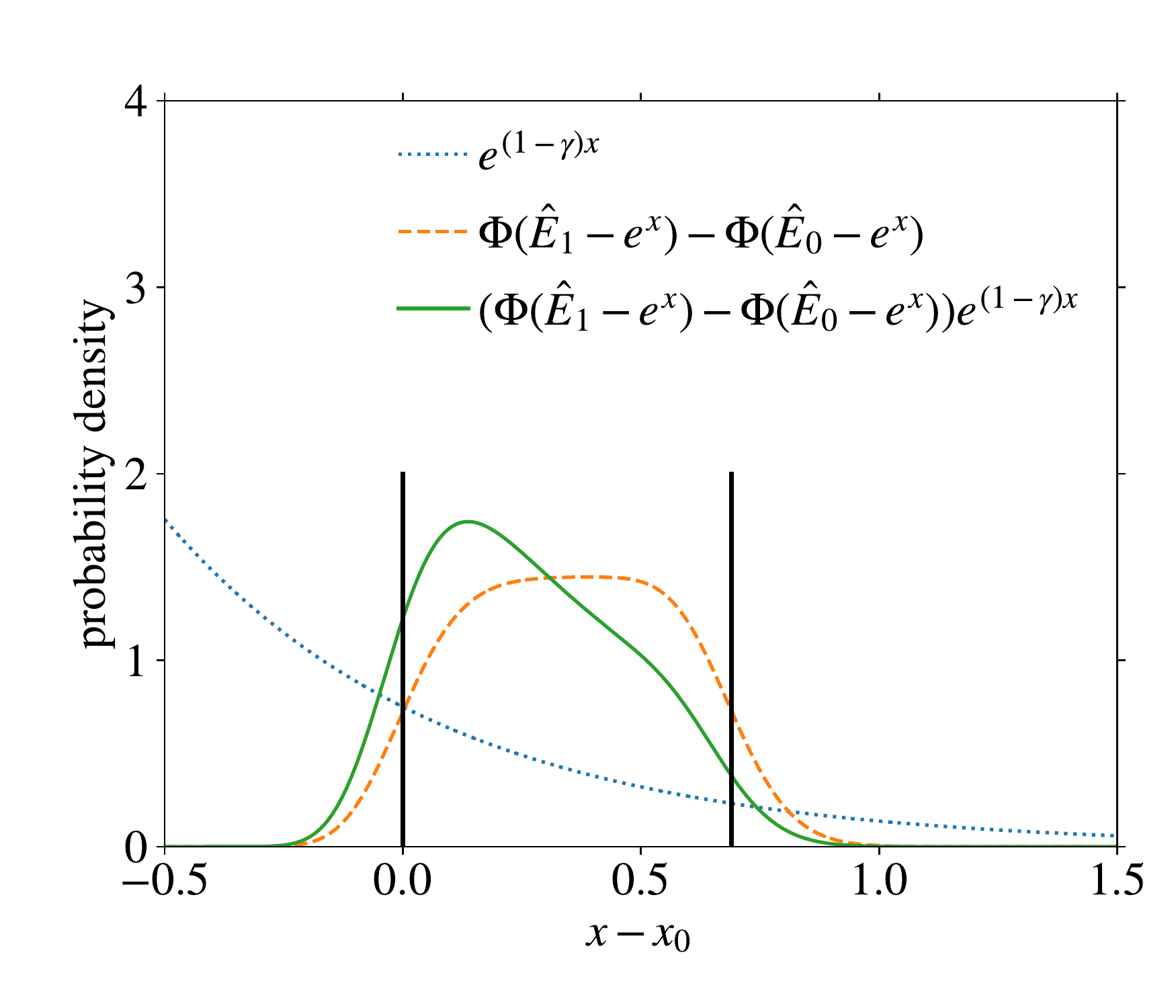}
\caption{Weighting function in the computation of $\mlnhn\!^\star$. Shown on the x-axis is the logarithm $x = \ln E$ of the true air shower energy $E$. Thick vertical lines indicate the boundaries of an energy bin with $\Delta \log_{10} E = 0.3$. A value of $\gamma = 2.7$ is used for energy spectrum, the energy resolution is $10\,\%$.}
\label{fig:binning}
\end{figure}

\begin{figure}
\includegraphics[width=\columnwidth]{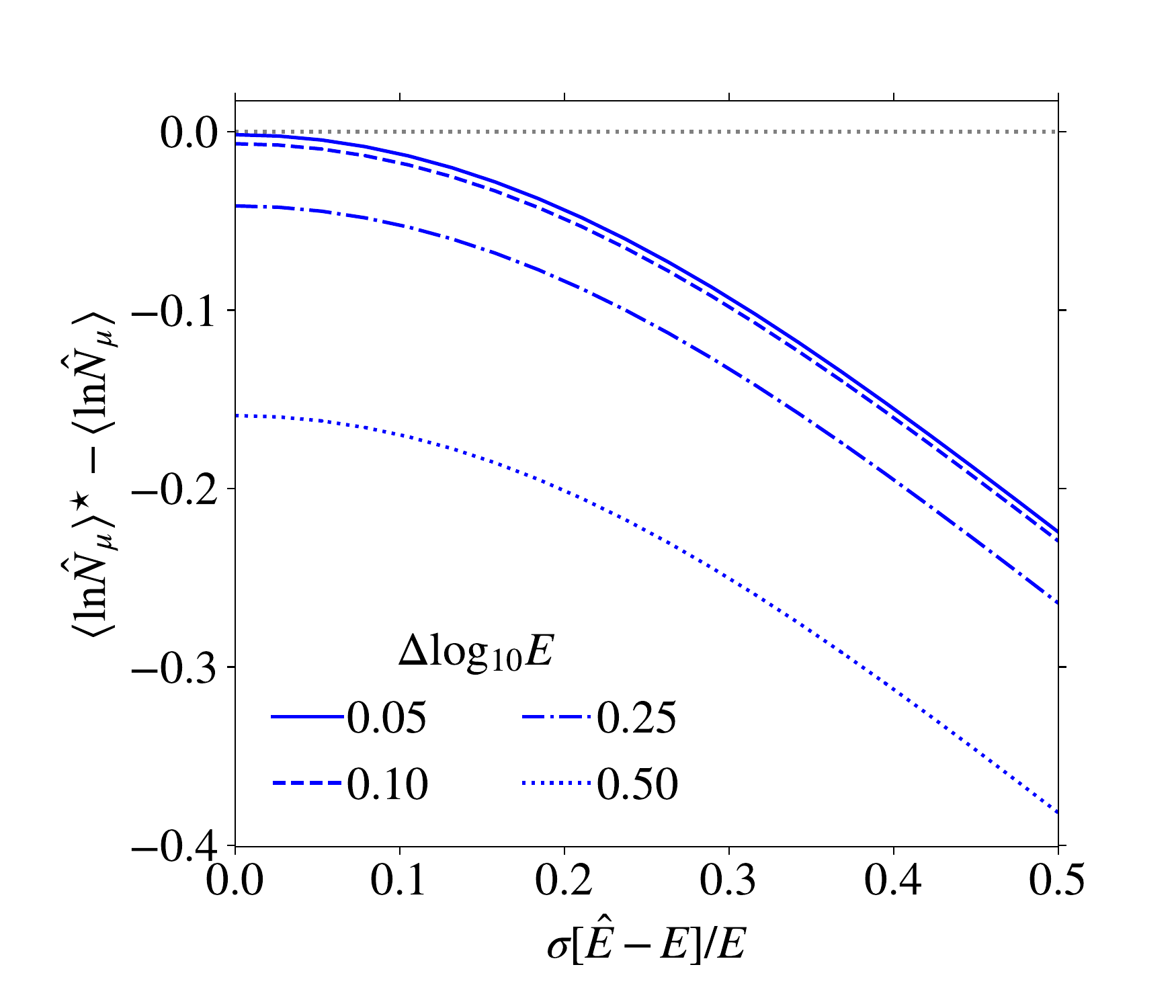}
\caption{Binning bias for a $E^{-2.7}$ spectrum and $\beta = 0.9$ as a function of the relative resolution of energy estimate $\hat E$.}
\label{fig:binning_bias}
\end{figure}

To compute $\mlnhn^\star$, we to convolve the integrand in \eq{binning_orig} with a energy resolution kernel. Usually, a normal distribution is appropriate
\begin{equation}
\norm(\hat E; E) = \frac 1 {\sqrt{2\pi}\sigma} e^{-\frac12(\hat E - E)^2/\sigma^2},
\end{equation}
which describes the probability to observe an energy estimate $\hat E$ with resolution $\sigma$ for a given true energy $E$. This leads to
\begin{gather}
\mlnhn\!^\star = \frac{\int_{\hat E_0}^{\hat E_1} \dd \hat E \int_{-\infty}^{\infty} \dd x\, \norm(\hat E; e^x) (\ln N_\mu^0 + \beta x) e^{(1 - \gamma) x}}{\int_{\hat E_0}^{\hat E_1} \dd \hat E \int_{-\infty}^{\infty} \dd x\, \norm(\hat E; e^x) e^{(1 - \gamma) x}} \nonumber\\
= \ln N_\mu^0 + \frac{\int_{-\infty}^{\infty} \dd x\, (\Phi(\hat E_1; e^x) - \Phi(\hat E_0; e^x)) \beta x e^{(1 - \gamma) x}}{\int_{-\infty}^{\infty}  \dd x\, (\Phi(\hat E_1; e^x) - \Phi(\hat E_0; e^x)) e^{(1 - \gamma) x}}.
\label{eq:mlnhn_bin_bias_eres}
\end{gather}
The integration over $\hat E$ was carried out in the second step, turning the probability density function $\norm(\hat E; e^x)$ into its cumulative density function $\Phi(\hat E; e^x)$. The weighting function for the integrand obtained in this way is illustrated in \fg{binning}. Showers with true energies near the lower edge of the bin get a higher weight and that also showers outside the bin interval contribute.

The effective energy interval to consider is now wider and not well bounded. \eq{mlnhn_bin_bias_eres} can be approximated by a Taylor series for $\Delta x \ll 1$ and $\sigma/E \ll 1$, but at this point it is easier to just compute the bias numerically by solving the equation. \fg{binning_bias} shows numerical solutions for several bin widths and energy resolutions.

The binning bias is comparable to the other biases previously considered. For a common bin width of 0.1 in $\log_{10}\hat E$, an energy resolution of 15\,\%, a spectral index $\gamma = 2.7$, and $\beta = 0.9$, the bias for $\mlnhn$ is $-0.03$. This translates into a bias of $-0.3$ for $\mlna$, about 7\,\% of the proton-iron distance. This bias is making the composition appear lighter.

\section{Conclusions}

The impact of stochastic fluctuations in the number of muons and the shower energy as well as in their experimental estimates on the computation of $\mlna$ was discussed. Only $\mlnn$ has a straight-forward relationship to the mean logarithmic mass $\mlna$ of cosmic rays. The biases calculated here are typically smaller than 10\,\% of the proton-iron distance, but can be larger for detectors with poor resolution. Several may need to be added.

To get the smallest systematic uncertainty, the muon number should be measured event-by-event and the mean logarithmic muon number $\mlnn$ computed, correcting for resolution and binning effects. A computation based on the mean muon number $\avg{N_\mu}$ is possible, but requires a correction that depends on the size of the natural fluctuations of $N_\mu$ for showers of the same energy, more precisely on $\var[(N_\mu - \avg{N_\mu}) / \avg{N_\mu}]$. This variance has to be measured or estimated from air shower simulations. If simulation results are reported, the variance $\var[(N_\mu - \avg{N_\mu}) / \avg{N_\mu}]$ should generally be included.

\section{Acknowledgments}

I am grateful for valuable discussions about this topic with Lorenzo Cazon and Felix Riehn. I also thank the reviewers from Astroparticle Physics who kindly commented on this draft and gave it more focus.

\bibliography{main}

\appendix

\section{Parametrization of relative variance of muon number}

\begin{figure}
\includegraphics[width=\columnwidth]{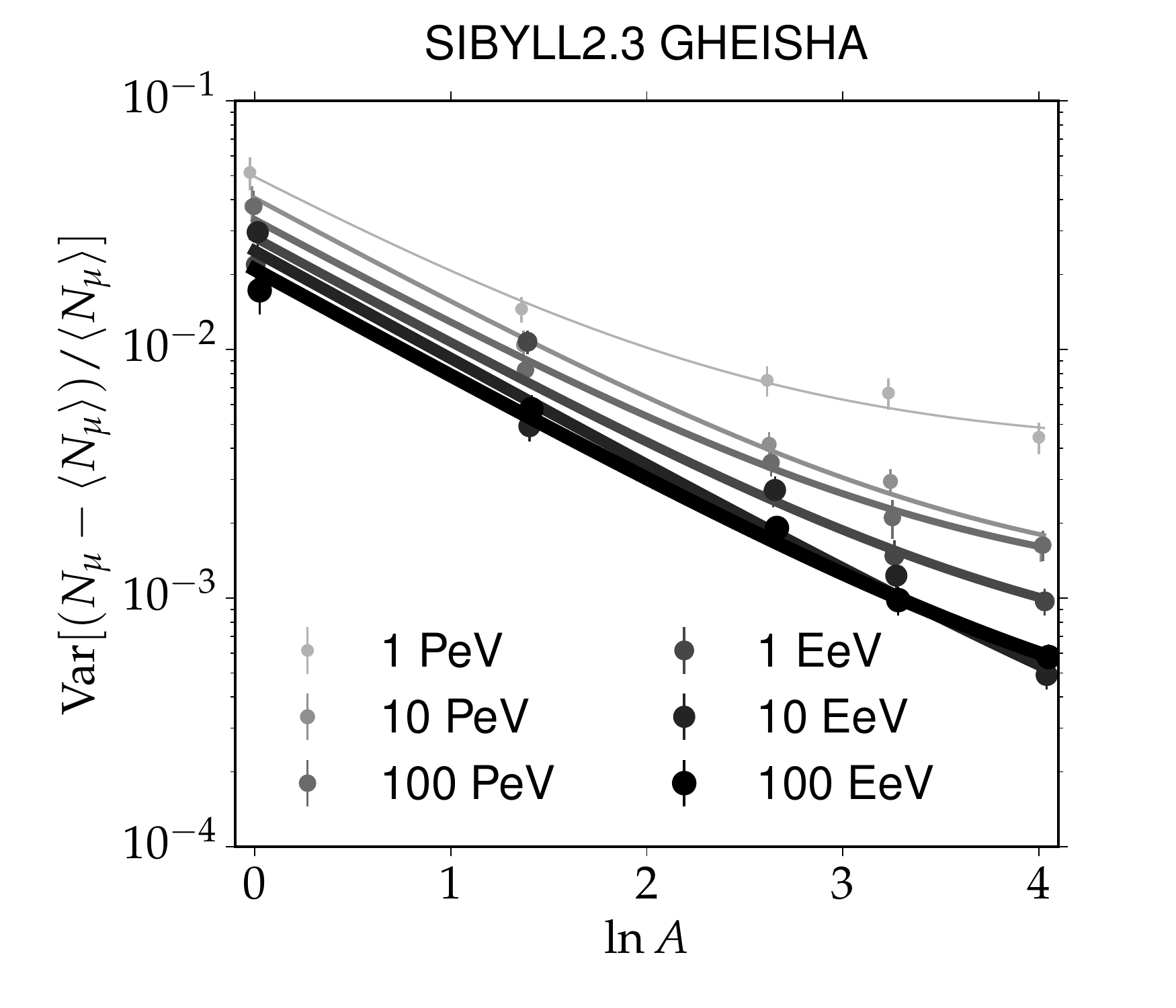}
\caption{Variance of the relative deviation from the mean muon number as a function of logarithmic mass $\ln A$ of the primary cosmic ray, using the data from \tb{nmu_sim}. Solid lines are fits described in the text.}
\label{fig:var_rel_nmu}
\end{figure}

The relative variance $\var[(N_\mu-\langle N_\mu \rangle)/ \langle N_\mu \rangle]$ of the muon number for primary cosmic rays with energy $E$ and mass $A$ plays an important role in section \ref{sec:lnmn_to_mlnn}. It is useful to have a parameterization for this quantity. Based on the numbers in \tb{nmu_sim}, the evolution is shown as a function of $\ln A$ for several energies in \fg{var_rel_nmu}. The simulations are well described by the model,
\begin{equation}\label{eq:var_parameters}
\var[(N_\mu - \langle N_\mu \rangle) / \langle N_\mu \rangle] = p_0(E) + p_1(E) / A,
\end{equation}
where $p_0$ and $p_1$ are energy-dependent parameters. The formula is motivated by the \emph{superposition model}~\cite{Matthews2005}, which states that an air shower with $A$ nucleons approximately behaves like a superposition of $A$ showers with an energy $E/A$. If the $A$ nucleons develop independently, the fluctuations in the individual sub-showers average out. This leads to a $1/A$ reduction in the variance. The other parameter $p_0$ summarizes correlated fluctuations which do not cancel, for example, fluctuations due to the depth of the first interaction.

\begin{figure}
\includegraphics[width=\columnwidth]{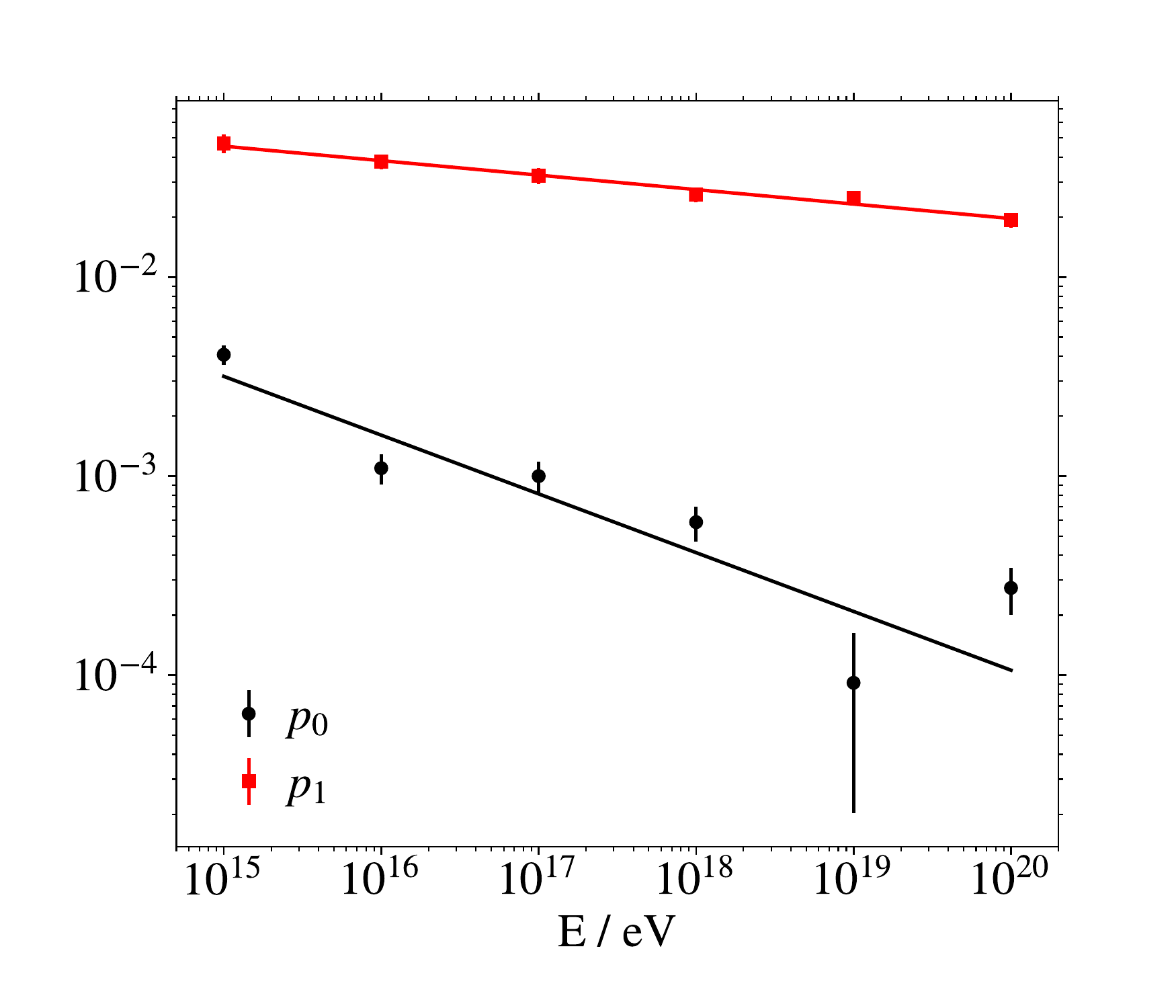}
\caption{Energy dependence of the fluctuation parameters from \eq{var_parameters}. Lines represent the fits described in the text.}
\label{fig:var_fit_parameters}
\end{figure}

The energy-dependence of the parameters $p_0$ and $p_1$ is shown in \fg{var_fit_parameters} and well described by a power law,
\begin{gather}
    p_i(E) = a_i \, (E/10^{15}\,\text{eV})^{b_i} \\
\begin{array}{l l}
    a_0 &= 0.00317 \pm 0.00037 \\
    a_1 &= 0.0455 \pm 0.0032
\end{array}\quad
\begin{array}{l l}
b_0 &= -0.295 \pm 0.033 \\
b_1 &= -0.0727 \pm 0.0095.
\end{array}
\end{gather}
These numerical values are valid for a set of pure primary cosmic rays with mass $A$ of vertical incidence, simulated with SIBYLL-2.3 in a  standard atmosphere. To compute the variance for mixtures of primaries, the mean $\langle N_\mu \rangle$ also needs to be parametrized as a function of $A$, which can be done with a power-law as well.

\section{Taylor series}

The following Taylor series are used in the paper:
\begin{gather}
\frac 1 {e^{(1-\gamma) x_0} x_0^{n} \Delta x} \int_{x_0}^{x_0 + \Delta x} e^{(1-\gamma) x} x^n \dd x \approx \nonumber \\
1 + \frac {\Delta x}{2} \left((1-\gamma) + \frac n {x_0}\right) \nonumber \\
+ \frac{\Delta x^2} 6 \left((1-\gamma)^2 + \frac{2n(1-\gamma)}{x_0^{n-1}} + \frac{n (n - 1)}{x_0^{n - 2}}\right) \nonumber \\
+ \bigo(\Delta x^3),
\label{eq:taylor1}
\end{gather}
\begin{gather}
\frac{a_0 + a_1 \Delta x + a_2 \Delta x^2}{b_0 + b_1 \Delta x + b_2 \Delta x^2} \approx \nonumber \\
\frac{a_0}{b_0} + x \frac{a_1 b_0 - a_0 b_1}{b_0^2} \nonumber \\
+ x^2 \frac{a_2 b_0^2 - a_1 b_0 b_1 + a_0 (b_1^2 - b_0 b_2)}{b_0^3} + \bigo(x^3)
\label{eq:taylor2}.
\end{gather}
Combining these two series, one gets
\begin{gather}
\frac{1}{x_0^n}\frac{\int_{x_0}^{x_0 + \Delta x} e^{(1-\gamma) x} x^n \dd x}{\int_{x_0}^{x_0 + \Delta x} e^{(1-\gamma) x} \dd x} \approx \nonumber \\
1 + \frac {\Delta x}{2} \frac n {x_0} + \frac {\Delta x^2}{6} \left(\frac{n (1-\gamma)}{2 x_0} + \frac{n (n - 1)}{x_0^2}\right) \nonumber \\
+ \bigo(\Delta x^3).
\label{eq:taylor3}
\end{gather}

\end{document}